\documentclass[superscriptaddress,twocolumn,pre]{revtex4}
\usepackage{amssymb}
\usepackage{natbib}
\usepackage{amsmath}
\usepackage{amsfonts}
\usepackage{graphicx}
\usepackage{subfigure}
\usepackage{mathrsfs}
\usepackage{dcolumn} 
\usepackage{bm}
\usepackage{color}
\usepackage{cancel}
\usepackage{mathbbol}
\usepackage{epsfig}
\usepackage{units}
\usepackage{esint}
\usepackage{soul} 
\usepackage{url}

\definecolor{rot}{rgb}{0.75,0.05,0.25}
\definecolor{hellgrau}{gray}{0.5}
\definecolor{blau}{rgb}{0,0,0.7}

\begin{document}

\title[]{Thermodynamics of a Quantum Annealer}

\author{Lorenzo Buffoni}
\email{lorenzo.buffoni@unifi.it}
\address{Department of Physics and Astronomy, University of Florence, Via Sansone 1, I-50019, Sesto Fiorentino (FI), Italy.}
\address{Department of Information Engineering, University of Florence, via S. Marta 3, I-50139 Florence, Italy}
\author{Michele Campisi}
\address{NEST, Istituto Nanoscienze-CNR and Scuola Normale Superiore, Piazza S. Silvestro 12, I-56127 Pisa,Italy}
\address{Department of Physics and Astronomy, University of Florence, Via Sansone 1, I-50019, Sesto Fiorentino (FI), Italy.}
\address{INFN Sezione di Firenze, via G. Sansone 1, I-50019, Sesto Fiorentino (FI), Italy.}

\begin{abstract} 
The D-wave processor is a partially controllable open quantum system that exchanges energy with its surrounding environment (in the form of heat) and with the external time dependent control fields (in the form of work). Despite being rarely thought as such, it is a thermodynamic machine. Here we investigate the properties of the D-Wave quantum annealers from a thermodynamical perspective. We performed a number of reverse-annealing experiments on the D-Wave 2000Q via the open access cloud server Leap, with the aim of understanding what type of thermal operation the machine performs, and quantifying the degree of dissipation that accompanies it, as well as the amount of heat and work that it exchanges. The latter is a challenging task in view of the fact that one can experimentally access only the overall energy change occurring in the processor, (which is the sum of heat and work it receives). However, recent results of non-equilibrium thermodynamics (namely, the fluctuation theorem and the thermodynamic uncertainty relations), allow to calculate lower bounds on the average entropy production (which quantifies the degree of dissipation) as well as the average heat and work exchanges. The analysis of the collected experimental data shows that 1) in a reverse annealing process the  D-Wave processor works as a thermal accelerator and 2) its evolution involves an increasing amount of dissipation with increasing transverse field. 
\end{abstract}

\maketitle

\section{Introduction}
In recent years, improvements in both size and controllability of quantum annealers \cite{kadowaki1998quantum,lanting2014entanglement} allowed to develop new and diverse applications. In particular the possibility of employing quantum annealers as quantum simulators \cite{king2018observation,harris2018phase,izquierdo2020testing} and quantum samplers \cite{benedetti2018quantum,benedetti2016estimation,vinci2019path,ayanzadeh2020reinforcement} have opened the possibility for any physicist to perform experiments without the need to own and maintain a lab. This opens as well to the possibility of experimentally exploring thermodynamic phenomena in the quantum regime \cite{deffnerthermo}. Gardas and Deffner \cite{gardas2018quantum}, for example, have investigated the (thermo)-dynamics of a D-Wave quantum annealer by using the Jarzynski equality \cite{jarzynski2007comparison} to quantify the degree by which its evolution deviates from an ideal unitary evolution. 
Here we take a step further and look at the entropy production, heat exchanged with the environment and work exchanged with the electronic control, thus gaining further understanding of the thermodynamics of quantum annealing. While one does not have experimental access to those quantities, our thermodynamical analysis allows to experimentally put bounds on them. The method can be applied to other quantum-thermodynamics experimental platforms as well.

In the following we first give a brief overview of quantum annealing, specifically of the D-Wave 2000Q processor. We then present general thermodynamic arguments that can be applied as well to any quantum platform, e.g., NMR systems and NV centres, \cite{Pal19PRA100,Hernandez-Gomez19arXiv190708240}.
Finally, we move on to the analysis of the data obtained from our experiments.

\section{The D-Wave quantum annealer}
The Hamiltonian that governs the evolution of D-Wave quantum annealers is that of a transverse field Ising model:
\begin{equation}
    H(s_t) = (1-s_t)\Gamma\sum _i \sigma ^x_i + s_t\left[ \sum _i h_i \sigma ^z_i + \sum _{<i,j>} J_{ij} \sigma ^z_i \sigma ^z_j\right]
\end{equation}
The physical system is composed of superconducting flux qubits arranged on a graph called ``Chimera''. The control of the effective local fields $h_i$ and qubit-qubit interactions $J_{ij}$ is achieved by controlling local magnetic fields generated by currents  circulating in coils on the chip. 
The latest processor, named D-Wave 2000Q has up to 2000 qubits each coupled with six neighbours. 

Note that the Hamiltonian is a weighted sum of two Hamiltonians $H_x=\Gamma\sum _i \sigma ^x_i$, and $H_z=  \sum _i h_i \sigma ^z_i +\sum J_{ij} \sigma ^z_i \sigma ^z_j$ with weights $1-s_t$ and $s_t$ respectively. We shall refer to $s$ as the annealing parameter. An annealing schedule is the specification of the function $t \to s_t$, in the time interval $[0,\tau]$.

In the standard quantum annealing schedule (which we will refer to as forward annealing)  $s$ ramps up linearly from $s=0$ to $s=1$ in the annealing time $\tau$: $s_t= t/\tau$, hence it changes the Hamiltonian from $H_x$ into $H_z$. The standard narrative is that when  the system is prepared in the ground state of $H_x$, for long enough annealing time $\tau$, the forward annealing takes the system adiabatically to the ground state of $H_z$. By measuring the system energy at the final time $\tau$, one is then able to experimentally obtain the minimum of $H_z$. Accordingly if an optimisation problem can be mapped onto the problem of finding the minimum of a function of the type $H_z$,
the annealer provides an experimental method to obtain its solution. That in short is the essence of quantum computing with quantum annealers.

\begin{figure}%
    \centering
    \includegraphics[width=7.5cm]{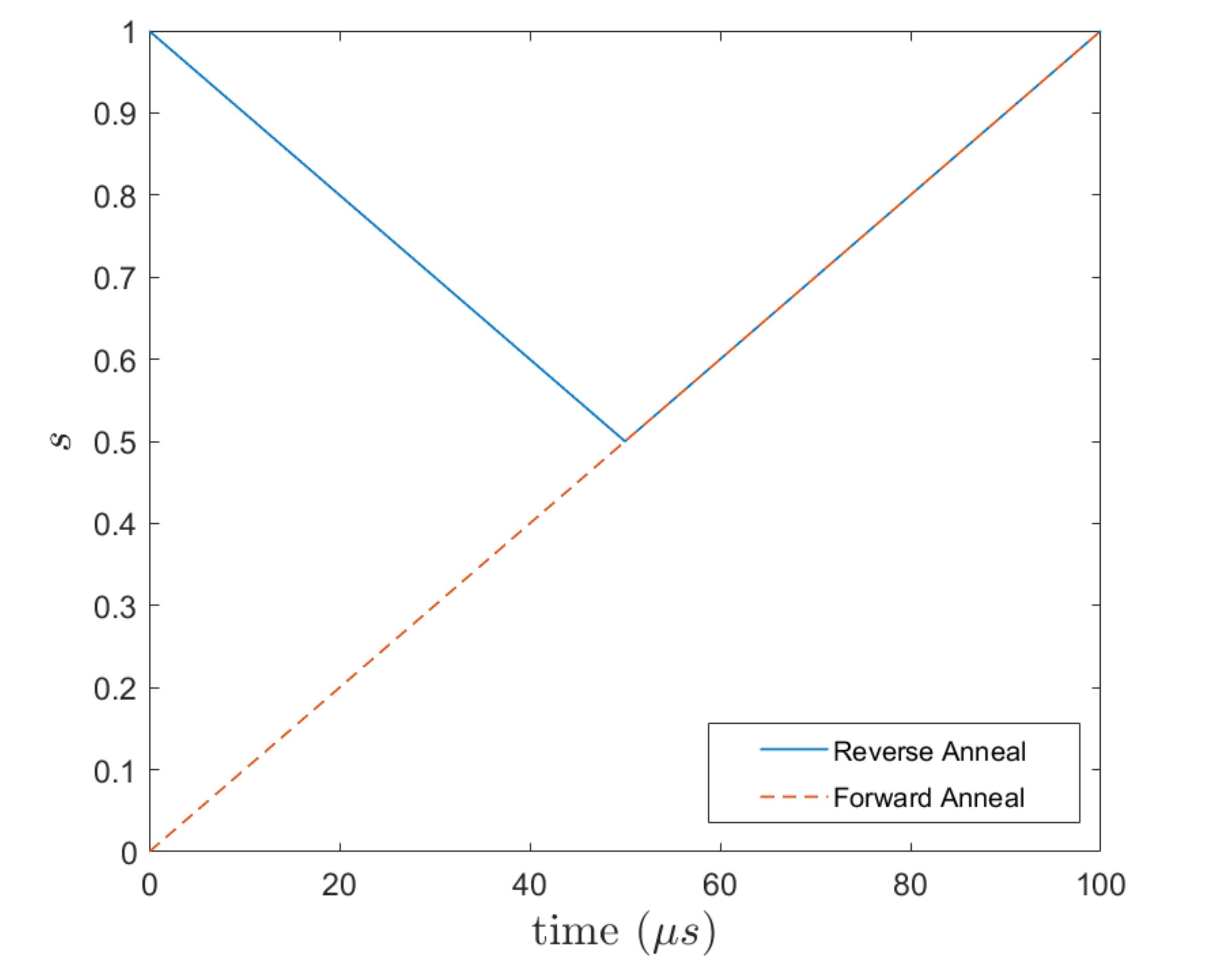} 
    \caption{Example of a reverse annealing schedule, Eq. (\ref{eq:reverseSchedule}), used in our experiments, see solid line. As a reference a forward schedule, see the dashed line, is also plotted.   
}
    \label{fig:schedule}%
\end{figure}

However evidence has been provided (see e.g. \cite{gardas2018quantum}) that the dynamics of the system is not unitary, namely despite all the efforts taken to effectively decouple the chip from external perturbations these inevitably disturb the system dynamics. According its dynamics are best described as that of an open quantum system. As we shall see below, our thermodynamical analysis corroborates those findings, and suggests that the environmental disturbance is not necessarily detrimental, as it in fact helps the system follow the ground state.

In this work we focus on the so called reverse annealing schedule:
\begin{align}
s_t = 
\left \lbrace \begin{array}{ll}
1-2(1-\bar s) t/\tau,  & t \in [0,\tau/2]  \\
-1+2\bar s + 2(1-\bar s)t/\tau, & t \in [\tau/2,\tau] 
\end{array}\right.
\label{eq:reverseSchedule}
\end{align}
where the annealing parameter $s$ starts at $s=1$, at time $t=0$, linearly decreases in time, until it reaches a minimum vale $\bar s$, at half annealing time $\tau/2$ and then goes back to $s=1$, with an ascending linear ramp ending at time $\tau$ (see Fig. \ref{fig:schedule}).

With this choice of annealing protocol we have the possibility to initialise the processor in any spin configuration. For the present study we generated a thermal sample of initial configurations, at inverse temperature $\beta_1$. This was achieved by preparing the system in a configuration $\{\bar \sigma_i^z\}$ with the relative frequency $e^{-\beta_1 E_z(\{\bar \sigma_i^z\})}/Z(\beta_1)$, given by the Boltzmann factor. Here $E_z(\{\sigma_i^z\})=\sum _i h_i \sigma ^z_i + \sum J_{ij} \sigma ^z_i  \sigma ^z_j$ is the energy of the configuration $(\{\sigma_i^z\})$, and $Z(\beta_1)= \sum e^{-\beta_1 E_z(\{\sigma_i^z\})} $, with the sum running over all possible configurations, is the according partition function.

Our experiments were performed on an antiferromagnetic chain of spins with length $l=300$ and no local fields, the Hamiltonian thus reads:
\begin{equation}
    H(s_t) = (1-s_t)\sum _i \sigma ^x_i + s_t \sum _{i} \sigma ^z_i \sigma ^z_{i+1}
    \label{eq:ourH}
\end{equation}
This choice is dictated by the fact that we know exactly its ground state (adjacent spins with opposite signs) which allows to 
easily probe how far the system gets away from it in the annealing time $\tau$.
Another reason for our particular choice is that the chain in Eq. (\ref{eq:ourH}), is a subgraph of the Chimera graph which can be easily implemented onto the D-Wave processor. If that were not the case one would have to use special techniques (i.e., the minor embedding technique \cite{choi2008minor}) that may potentially affect the accuracy of the experimental results.

\section{Thermodynamics}
\begin{figure}%
    \centering
    \includegraphics[width=0.5\linewidth]{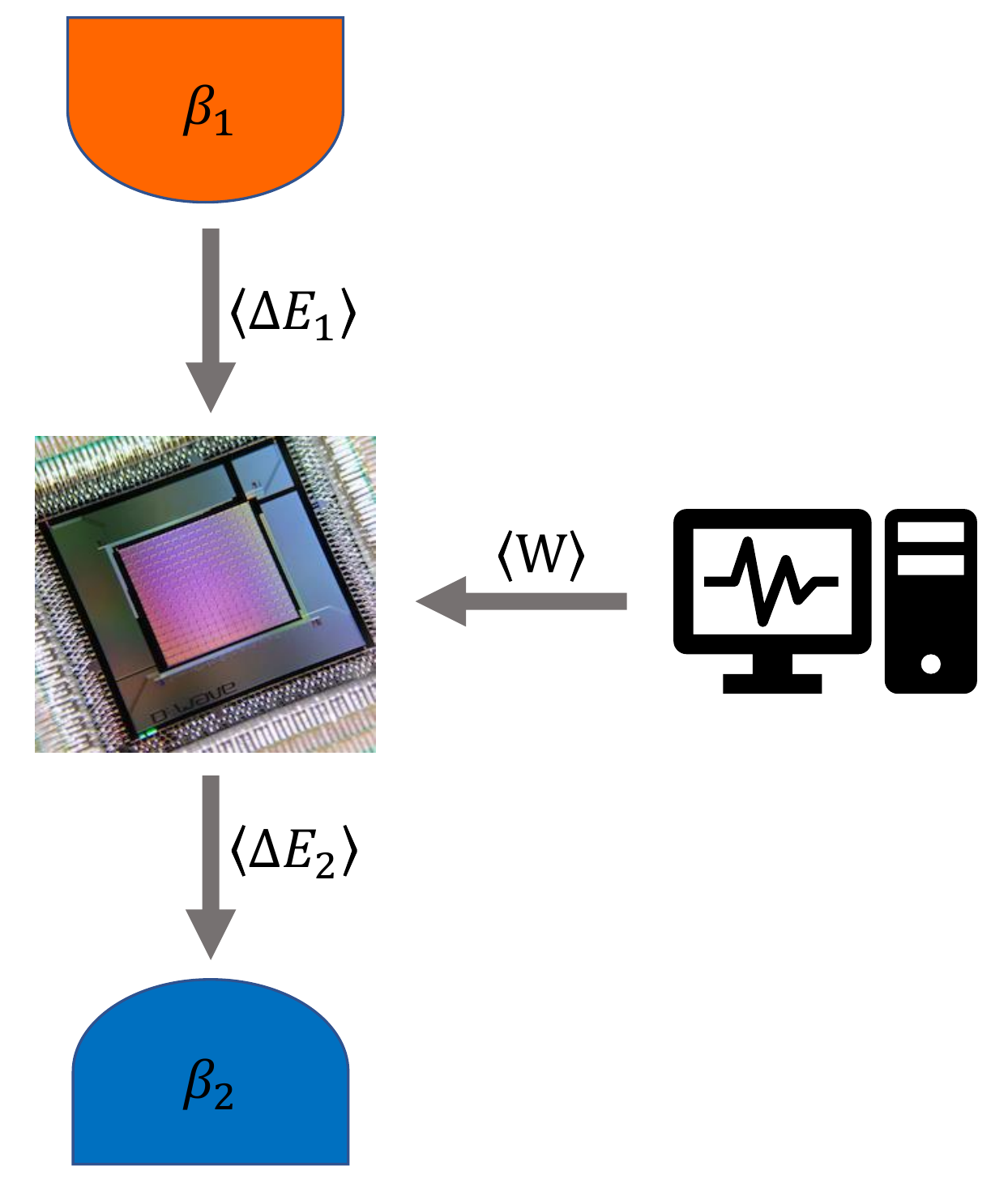} 
    \caption{Schematic representation of the D-Wave processor as a thermal engine.}
    \label{fig:engine}%
\end{figure}

The D-wave processor is a driven open quantum system. Namely, it is a physical system that interacts both electromagnetically with external control fields (with which it exchanges work), and thermally with a thermal environment at very low temperature $T_2$, 
with which it exchanges heat. Despite being rarely thought as such, it is a thermodynamic machine, see Fig. (\ref{fig:engine}).

As mentioned above, in this work we consider the situation where the processor is prepared at some temperature $T_1 > T_2$, and undergoes an annealing process, described by some schedule $s_t$ of the annealing parameter of time duration $\tau$.
Accordingly, the system$+$environment compound may be seen as an isolated driven bipartite system starting in the direct product state:
\begin{align}
\rho = \frac{e^{-\beta_1 H_1}}{Z_1} \otimes \frac{e^{-\beta_2 H_2}}{Z_2}
\end{align}
where $\beta_i= (k T_i)^{-1}$, $k$ is Boltzmann's constant, and $H_{1,2}$ are respectively the processor's Hamiltonian at time $t=0$, and the environment Hamiltonian. 
As such it obeys the multivariate fluctuation theorem, which in the specific case of a cyclic schedule (i.e.., such that $s_0=s_\tau$, as is the case of reverse annealing) reads  \cite{Campisi15NJP17}:
\begin{align}
\frac{p(\Delta E_1,\Delta E_2)}{p(-\Delta E_1,-\Delta E_2)}= e^{\beta_1 \Delta E_1 +\beta_2 \Delta E_2}
\label{eq:MVFT}
\end{align}
where $\Delta E_i, i=1,2$ are, respectively, the (stochastic) energy changes of the processor and its environment, occurring in the schedule time $\tau$ (here we assume the so called ``two-point-measurement" scheme \cite{Campisi11RMP83,Campisi11RMP83b}) and $p(\Delta E_1,\Delta E_2)$ is the joint probability of their occurrence in a single run of the reverse annealing schedule.
The multivariate fluctuation theorem above has a number of important consequences. First of all, it implies \cite{Campisi15NJP17}
\begin{align}
 \left< \Sigma \right>  \doteq \beta_1\left<\Delta E_1\right> + \beta_2\left<\Delta E_2\right> \geq 0 \, .
    \label{eq:2ndlaw}
\end{align}
By identifying $-\left<\Delta E_2\right>$ with the average heat provided to the processor by the cold bath during the annealing schedule, and $-\left<\Delta E_1\right>$ the average heat provided during the preparation of its initial hot state one can recognise the above inequality as Clausius inequality \cite{Fermi56Book}, which expresses the second law of thermodynamics. It is worth remarking that in our case there is only one physical bath at temperature $T_2$, but since we prepare the system at temperature $T_1$, it is as 
if there were another bath at temperature $T_1$, which interacted with our system between one annealing run and the next one. The quantity $\left<\Delta E_1\right>$ would then represent the average energy that that virtual bath received.

We note that $\left<\Delta E_1\right> + \left<\Delta E_2\right>$ represents the average work 
performed on the system+environment compound by the external driving:
\begin{align}
 \left< W \right> = \left<\Delta E_1\right> + \left<\Delta E_2\right> 
    \label{eq:1stlaw}
\end{align}

As shown in Ref. \cite{Solfanelli19arXiv} the combination of Eq. (\ref{eq:2ndlaw}) and (\ref{eq:1stlaw}), with the convention $0<\beta_1<\beta_2$ (i.e. $T_1>T_2>0$) is compatible with only four combinations of signs for $\left<\Delta E_1\right>,\left<\Delta E_2\right>,\left< W \right>$, each of which identifies an allowed thermal operation:
\begin{align}
\begin{array}{llll}
$[R]$: & \langle \Delta E_1 \rangle \geq 0 & \langle \Delta E_2 \rangle \leq  0 & \langle W \rangle \geq 0 \\
$[E]$: & \langle \Delta E_1 \rangle \leq 0  & \langle \Delta E_2 \rangle \geq 0 & \langle W \rangle \leq  0 \\
$[A]$: & \langle \Delta E_1 \rangle \leq 0  & \langle \Delta E_2 \rangle \geq 0 & \langle W \rangle \geq 0 \\
$[H]$: & \langle \Delta E_1 \rangle \geq 0  & \langle \Delta E_2 \rangle \geq 0 & \langle W \rangle \geq 0 \, .
\label{eq:modes}
\end{array}
\end{align}
They correspond to [R] Refrigerator: heat flows from the cold bath to hot bath, with energy injection from the external driving; [E] energy Extraction (heat engine): part of the energy naturally flowing from the hot bath to the cold bath is derouted towards the driving apparatus; [A] thermal Accelerator: the driving provides energy to facilitate the natural flow from the hot bath to the cold bath; [H] Heater: both baths receive energy from the external driving. 

One of the aims of the present work is to single out which out of the four thermal operations occurs in a typical reverse annealing schedule. As we shall see, our experiments show that the D-Wave processor operates as a thermal accelerator, as represented in Fig. (\ref{fig:engine}).

Another aim of our work is to quantify the entropy production $\left<\Sigma\right>$, Eq. (\ref{eq:2ndlaw}), the work $\left< W\right>$, and the heat exchanged with the environment $\left< Q \right>\doteq- \left< \Delta E_2\right>$. This is a challenging task, because the hardware allows only to experimentally access the processor energy change $\left< \Delta E_1\right>$ which is the sum of work and heat $\left< \Delta E_1\right>=\left< Q \right>+\left< W\right>$. This is a typical situation encountered as well in other platforms, e.g. NV centres \cite{Hernandez-Gomez19arXiv190708240}.To partially solve the problem we invoke a general result that has been proved recently, known as thermodynamic uncertainty relation (TUR)\cite{Barato15PRL114,Hasegawa19PRL123,Timpanaro19PRL123,Zhang19arXiv}.
If a joint probability distribution $p(\sigma, \phi)$ obeys the fluctuation relation:
\begin{align}
\frac{p(\sigma, \phi)}{p(-\sigma, -\phi)}=e^\sigma \,
\end{align}
then \cite{Timpanaro19PRL123,Zhang19arXiv}
\begin{align}
\langle \phi \rangle^2 \leq \langle \phi^2 \rangle f(h^{-1}(\langle \sigma \rangle))
\label{eq:TUR}
\end{align}
where $f(x)=\tanh^2(x/2)$, and $h^{-1}$ is the inverse of $h(x)=x\tanh(x/2)$. After some manipulations, Eq. (\ref{eq:TUR}) can be rewritten as a bound on $\langle \sigma \rangle$:
\begin{equation}
    \left<\sigma\right> \geq 2 g \left(\frac{\langle \phi \rangle}{\sqrt{\left<\phi^2 \right>} }\right)
\end{equation}
where
$g(x)= x\tanh^{-1}(x)$.

By looking at Eq. (\ref{eq:MVFT}) we see that by exchanging the variable $\Delta E_2$ for the new variable $\Sigma = \beta_1 \Delta E_1+\beta_2\Delta E_2$, it is
\begin{align}
\frac{p(\Sigma, \Delta E_1)}{p(-\Sigma, -\Delta E_1)}=e^\Sigma \, .
\end{align}
Accordingly,
\begin{equation}
    \left<\Sigma\right> \geq 2 g\left(\frac{\langle \Delta E_1 \rangle}{\sqrt{\left<\Delta E_1^2 \right>} }\right)\, .
    \label{eq:SigmaBound}
\end{equation}
Note that the function $g$ is even, non-negative and gets the value $0$ only at $x=0$, accordingly Eq. (\ref{eq:SigmaBound}) implies Eq. (\ref{eq:2ndlaw}) \footnote{It also shows that in order to have no dissipation dissipation, the ratio ${\langle \Delta E_1 \rangle}/{\sqrt{\left<\Delta E_1^2 \right>} }$ must be made null, which can either be achieved at vanishing $\langle \Delta E_1 \rangle$ or diverging $\left<\Delta E_1^2 \right>$}.

Combined with Eqs. (\ref{eq:2ndlaw},\ref{eq:1stlaw}), Eq. (\ref{eq:SigmaBound}) gives as well bounds on heat dumped into the environment and work performed on the system+environment:
\begin{align}
-\left<Q \right> & \geq \frac{2}{\beta_2} g\left(\frac{\langle \Delta E_1 \rangle}{\sqrt{\left<\Delta E_1^2 \right>} }\right)
-\frac{\beta_1}{\beta_2}\left<\Delta E_1 \right>
\label{eq:Qbound}\\
\left<W \right> & \geq  \frac{2}{\beta_2} g\left(\frac{\langle \Delta E_1 \rangle}{\sqrt{\left<\Delta E_1^2 \right>} }\right)
+\left(1-\frac{\beta_1}{\beta_2} \right)\left<\Delta E_1 \right>\, .
\label{eq:Wbound}
\end{align}
Accordingly, by preparing the system at a known temperature $T_1$, and estimating the temperature, $T_2$, of the environment, the statistics of $\Delta E_1$ provides bounds on the average heat and work. Therefore, while entropy production, work and heat cannot be accessed experimentally, we can experimentally determine lower bounds on their average values by collecting the statistics of the energy change $\Delta E_1$ of the processor. Knowing the bounds may be sufficient to gaining useful information, such as which operation mode in Eq. (\ref{eq:modes}) occurs in a device. As we shall see, that is the case of our experiments.

Our work has therefore a two-fold value. On one hand it provides a quantitative understanding of the thermodynamics of quantum annealing.
On the other it illustrates how powerful the fluctuation relations can be in providing information that cannot be directly accessed experimentally, thus providing a general method that is applicable to any quantum-thermodynamics experimental platform. 

We remark that, at variance with other experimental works in the field, rather than aiming at experimentally verifying known theoretical results, here we use them to earn information that would be otherwise unavailable.

\section{Results}

\begin{figure}%
    \centering
    \includegraphics[width=7.5cm]{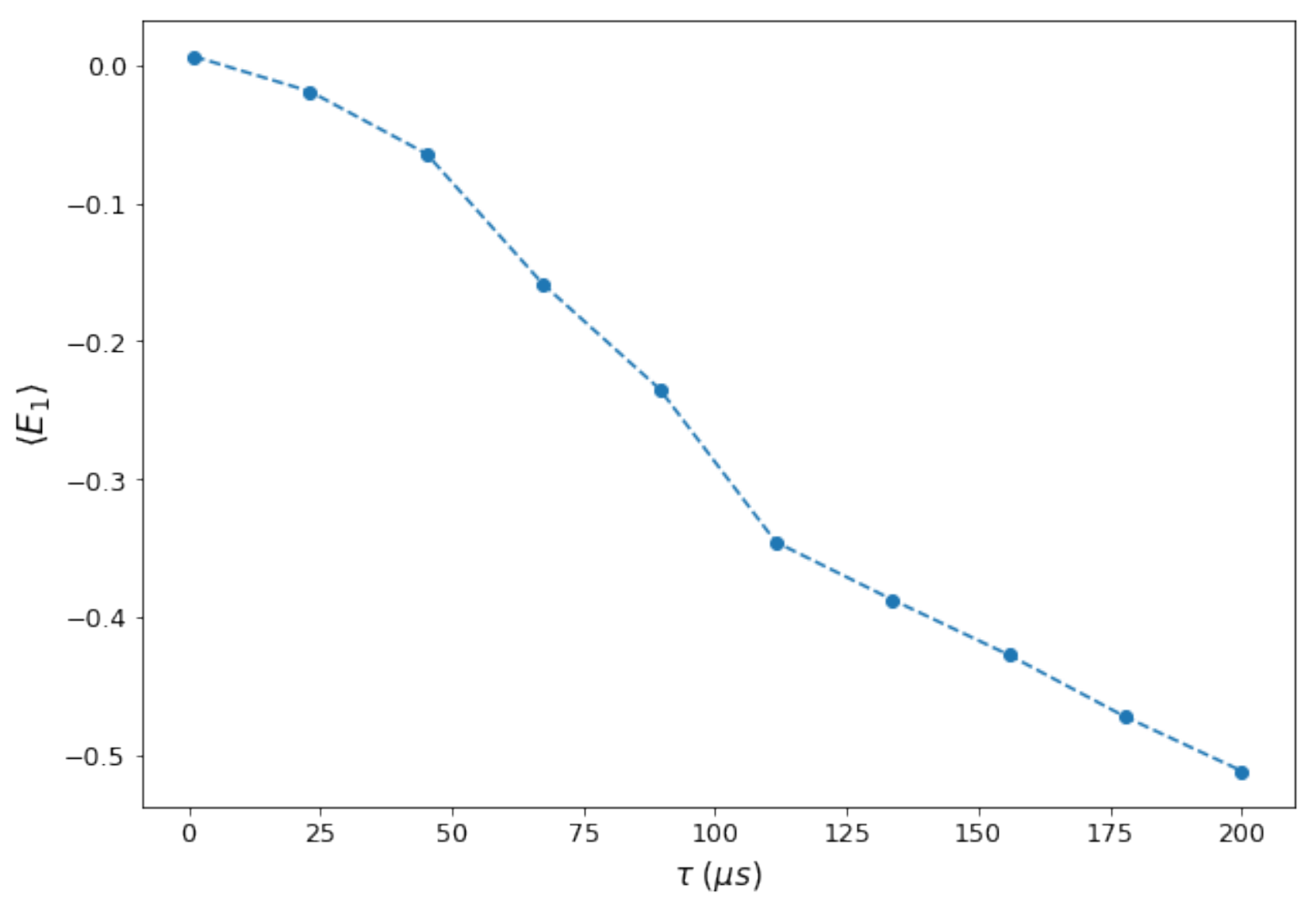} 
    \caption{Average energy per spin as a function of time for the constant schedule $s_t=1$. The initial average energy per spin is $0$, corresponding to the infinite temperature preparation. The average energy per spin of the ground state is $-2$
    }
    \label{fig:thermal}%
\end{figure}

We first studied how the average energy of the processor changes in the time interval $[0,\tau]$ in absence of external driving, namely for the schedule $s_t= 1, t \in [0,\tau]$, for various values of $\tau$ ranging from $\tau=0$ to $\tau= 200 \mu s$.
The results are presented in Fig. \ref{fig:thermal}. They are obtained by initialising the processor in a random configuration $\{\bar \sigma_i^z\}$ from a thermal distribution at $\beta_1=0$, namely each and all configurations have the same probability, regardless of their energy. The processor is then let evolve freely until time $\tau$, when we read the according final energy. For each repetition of the protocol we can accordingly record the stochastic energy change $\Delta E_1$ and with a sufficiently large number of repetitions build its statistics.  For each fixed value of $\tau$ we initialised the processor with 1000 different initial random spin configurations and for each such configuration we repeat the schedule $s_t$ ten times. At time $t=\tau$ we read the processor energy. Accordingly each datapoint in our graphs collect information from $10^4$ samples. 

As Fig. \ref{fig:thermal} shows, our system looses energy into the environment. From the thermodynamical point of view the process that is occurring is a spontaneous heat flow from a hot body (our virtual bath at $T_1=\infty$) to a colder one, namely the processor's thermal environment. It  evidences that the system cannot be  considered at all as an isolated quantum system. On the contrary, it is an open system that is exchanging energy. Since there is no external driving, this is only possible provided its dynamics is not unitary.

Inspecting the shape of the decay in Fig. \ref{fig:thermal} reveals no hint of an exponential form. This suggests that there are multiple times scales at play during the decay. Note that the end-scale of $200\mu s$ (beyond which experiments are currently not allowed), roughly marks the time at which $1/4$ of the energy of the chip has been given to the environment. 

\begin{figure}%
    \centering
    \begin{subfigure}{}
        \includegraphics[width=7.5cm]{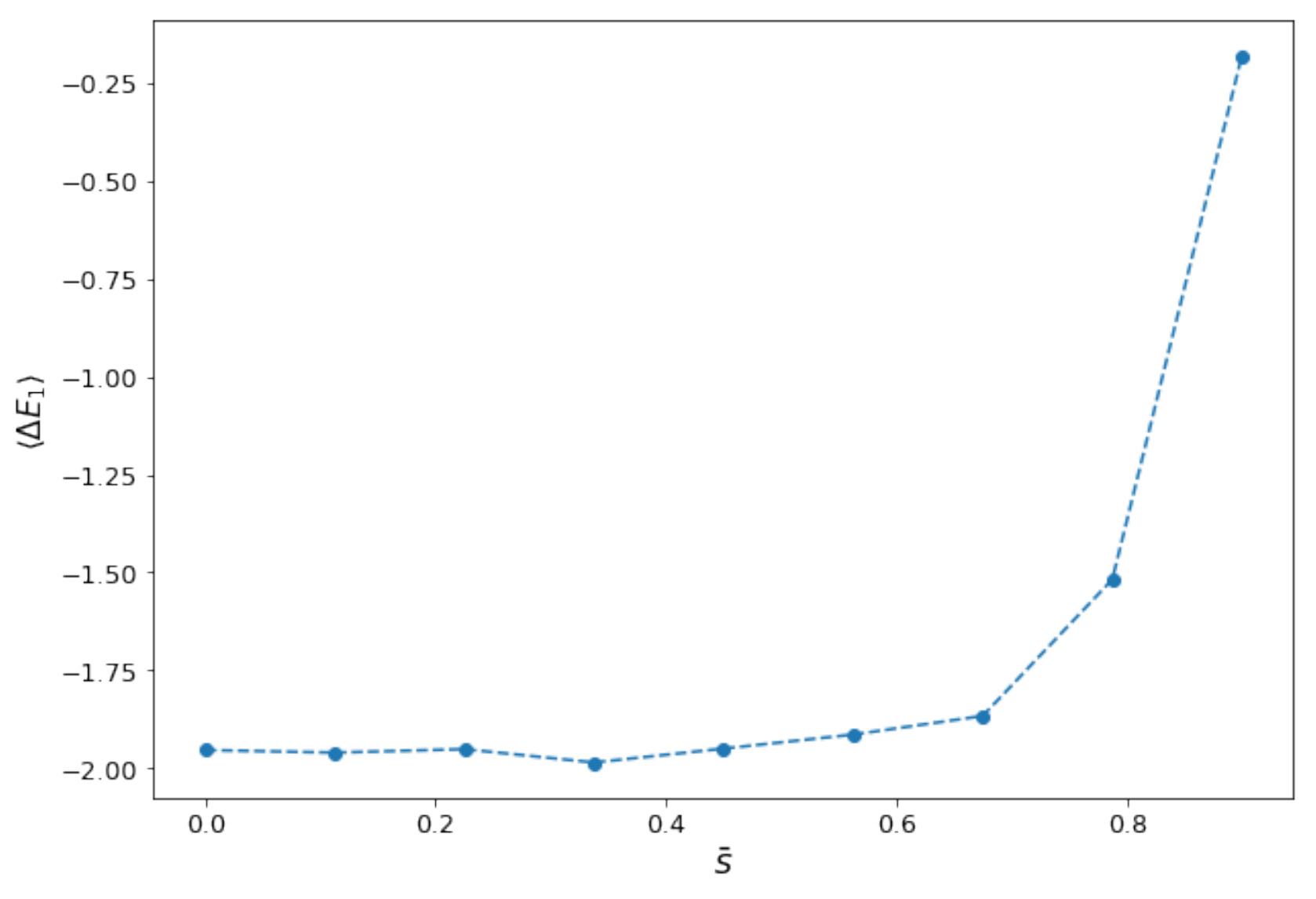}
    \end{subfigure}
    \begin{subfigure}{}
        \includegraphics[width=7.5cm]{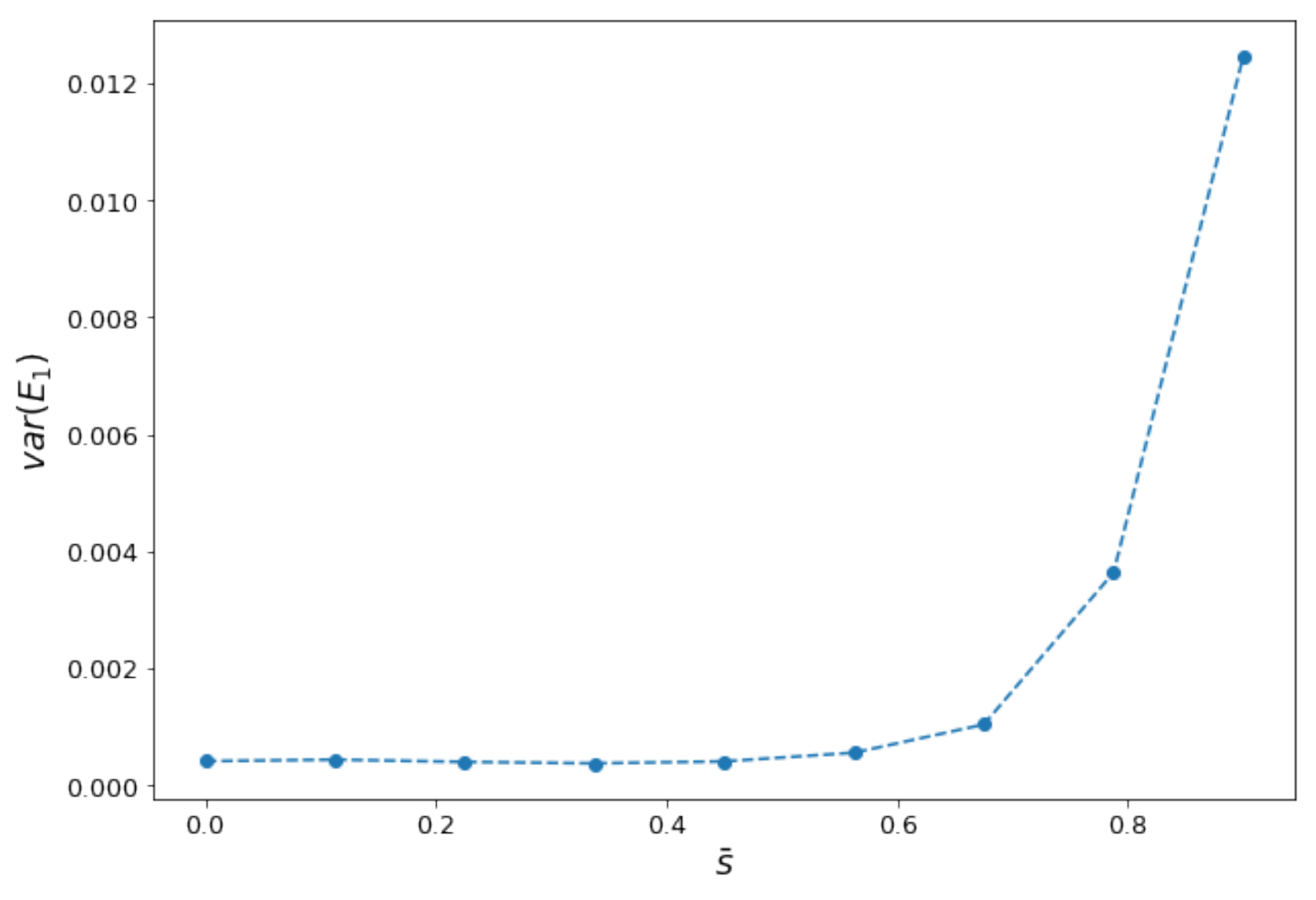}
    \end{subfigure}
    \caption{Average final energy per spin (panel a) and variance of final energy per spin (panel b) functions of $\bar{s}$, see Eq(\ref{eq:reverseSchedule}), for a fixed annealing time of $\tau = 100 \mu s$. }    
    \label{fig:distribution}%
\end{figure}

We then moved to investigate the energy exchanges during reverse annealing, as a function of the minimal annealing parameter $\bar s$, for a fixed annealing time $\tau= 100 \mu s$, and $\beta_1=0$.

Fig. \ref{fig:distribution} reports plots of the average energy change of the processor $\langle \Delta E_1\rangle$, and of the variance of its final energy distribution $var(E_{1})$, namely $\langle E_{1,f}^2\rangle- \langle E_{1,f}\rangle^2 $.
The energies are normalised by the chain length, so the ground state has energy $E=-2$. On the x-axis are the various values of $\bar{s}$ in the reverse annealing protocols. We notice that both the mean and the variance of the energy distribution exhibit a sharp decrease as $\bar{s}$ goes down, reaching a plateau below the value $1/2$. Thus albeit starting from a flat distribution with zero mean (infinite temperature distribution) the processor ends up with a final energy distribution that is remarkably close and peaked around the ground state of the system, when $\bar{s}< 1/2$.

\begin{figure}%
    \centering
    \begin{subfigure}{}
        \includegraphics[width=4.5cm]{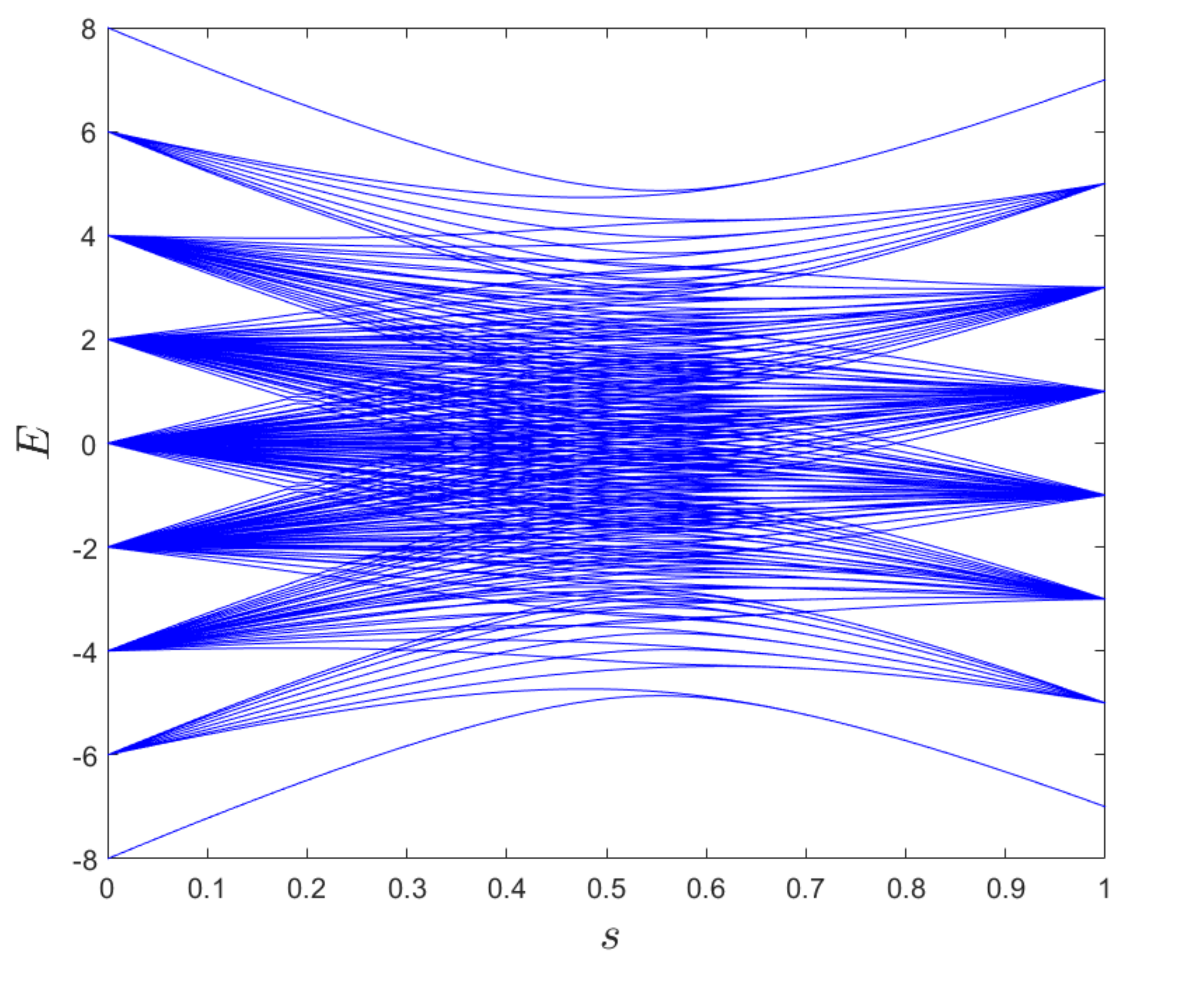}
    \end{subfigure}
    \begin{subfigure}{}
        \includegraphics[width=4.5cm]{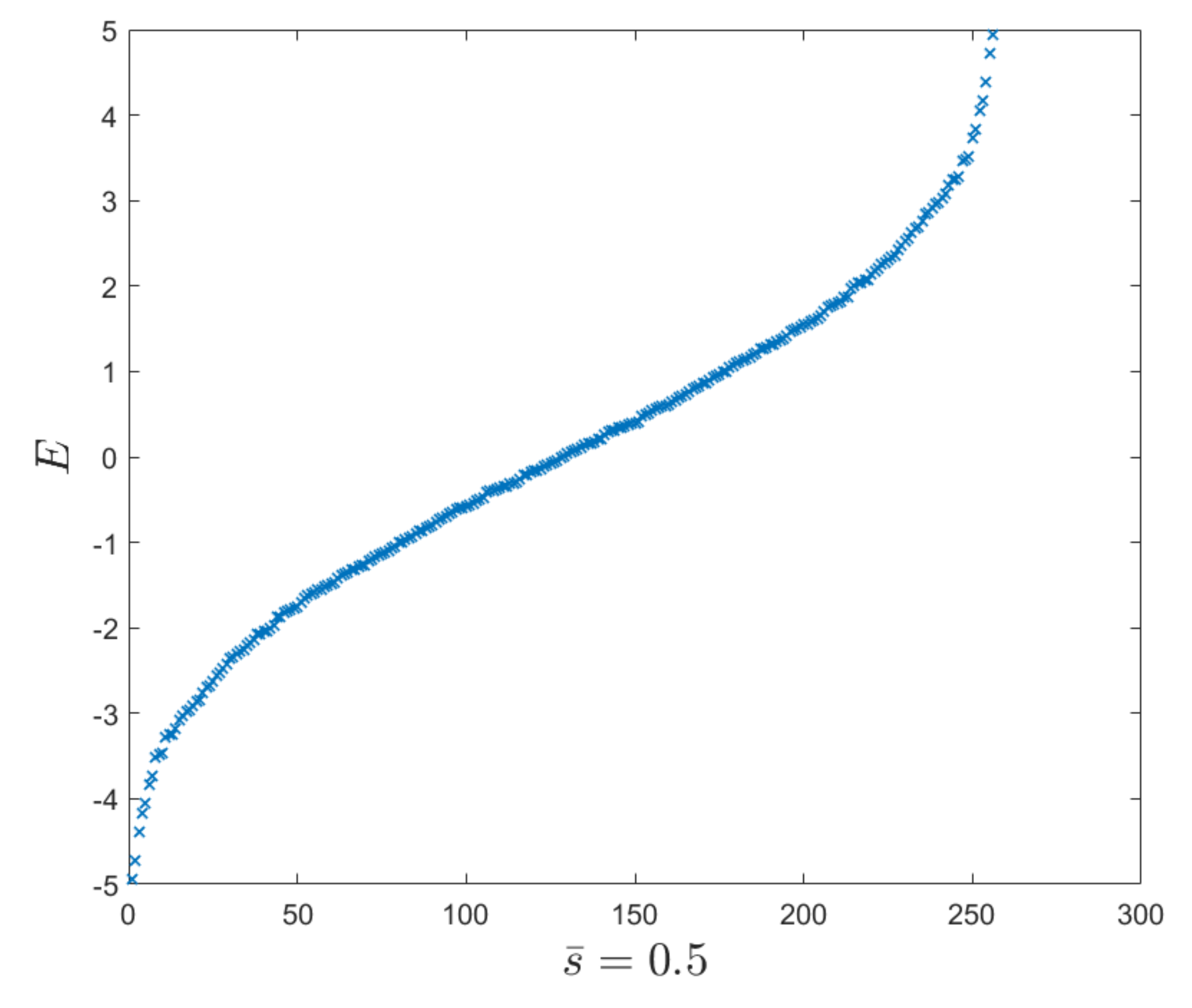}
    \end{subfigure}
    \begin{subfigure}{}
        \includegraphics[width=4.5cm]{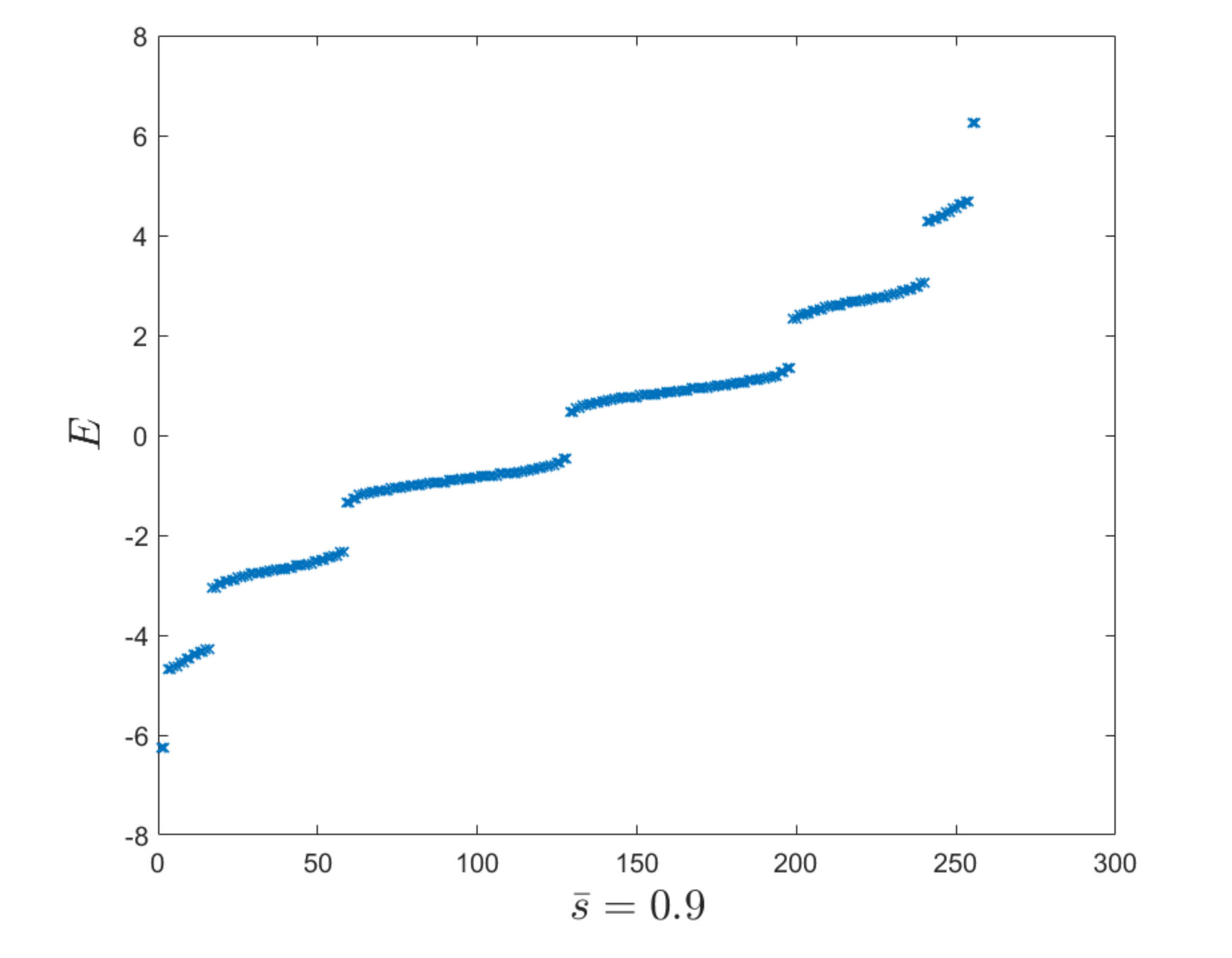}
    \end{subfigure}
    \caption{Panel a): Spectrum of $H(s)$, Eq. (\ref{eq:ourH}) as a function of annealing parameter $s$. Panel b): Spectrum of $H(s)$ at $s=0.5$. Panel c) Spectrum of $H(s)$ at $s=0.9$. These plots are for a chain of length 8.
    }
    \label{fig:spectrum}%
\end{figure}

In order to gain insight into the physics underling this behaviour, we investigated the energy spectrum of the Hamiltonian as a function $s$, see Fig. \ref{fig:spectrum}. The spectrum presents gaps which are largest at $s=0,1$ (where it presents multiple degeneracies) and shrink as $s=1/2$ is approached, where they get their minimal value. Accordingly, as $s=1/2$ is approached, the processor presents more and more frequencies that can resonantly couple to the frequencies of its environment. In short when getting close to $s=1/2$ more and more channels of interaction with the environment become available and the system becomes more and more prone to environmental effects. 

We remark that, in the thermodynamic limit, the system undergoes a quantum phase transition at $s=1/2$. In this regard it is worth mentioning the studies in Refs. \cite{gardas2018defects} regarding the crossing of the quantum critical point, and the according generation of excitations as predicted by the Kibble-Zurek mechanism \cite{zurek2005dynamics,dziarmaga2005dynamics,rossini2020dynamic,deffner2017kibble}.

\begin{figure}%
    \centering
    \includegraphics[width=7.5 cm]{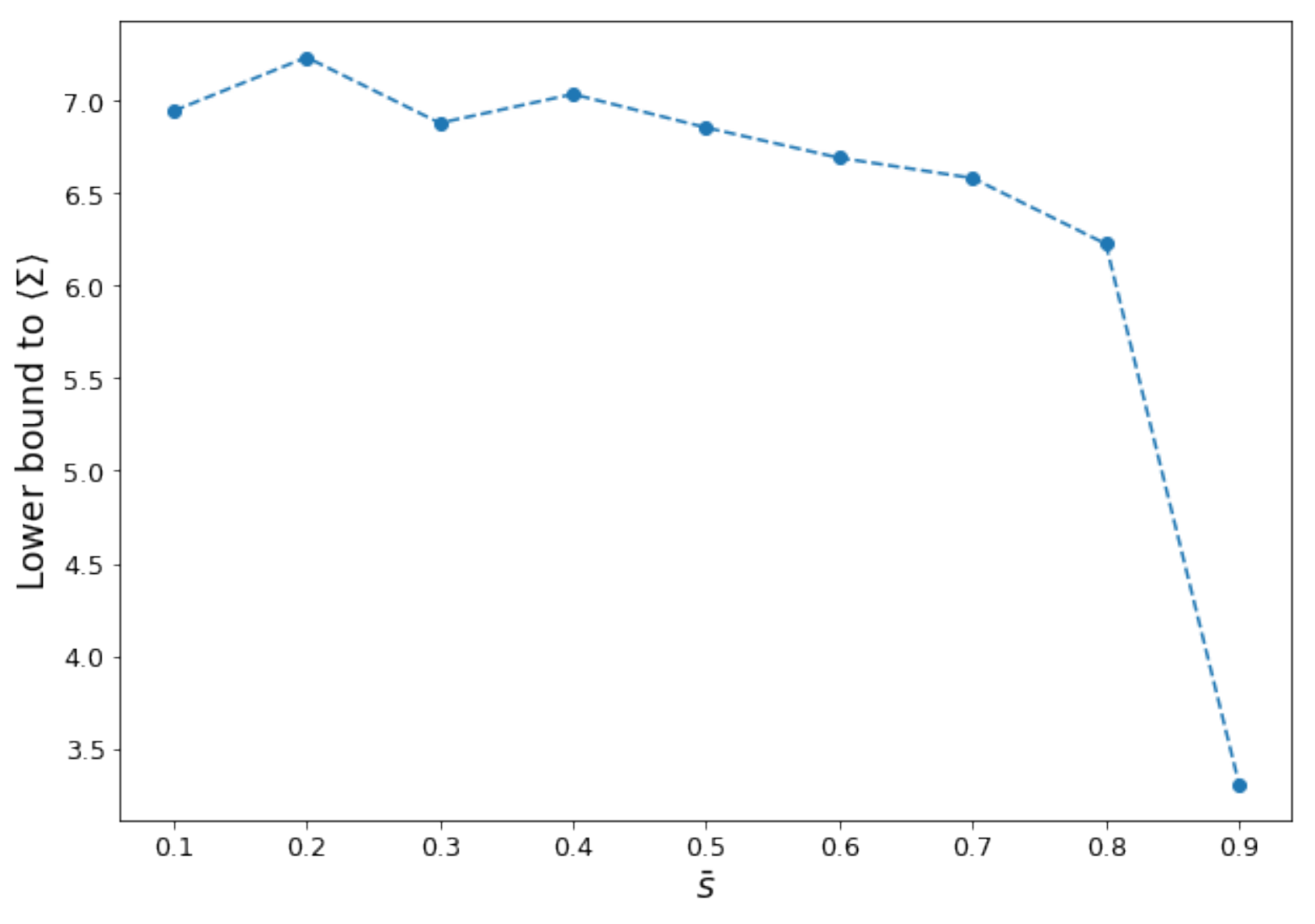} 
    \caption{Experimentally determined lower bound on average irreversible entropy production as a function of $\bar{s}$, Eq. (\ref{eq:SigmaBound}).
    }
    \label{fig:entropy}%
\end{figure}

Figure \ref{fig:entropy}  shows results pertaining to the lower bound to the irreversible entropy production, as calculated form Eq. (\ref{eq:SigmaBound}). 
The plot suggests that entropy production gets its highest values for $\bar s<1/2$: when $\bar s <1/2$ the interaction with the environment is most effective, more heat is dissipated in the environment, and more entropy is produced accordingly.

In the case $T_1=\infty$ ($\beta_1=0$) Eqs. (\ref{eq:Qbound},\ref{eq:Wbound}) get the simpler form:
\begin{align}
-\left<Q \right> & \geq  \frac{2}{\beta_2} g\left(\frac{\langle \Delta E_1 \rangle}{\sqrt{\text{var}(\Delta E_1)}} \right)
\label{eq:Qbound-simple}\\
\left<W \right> & \geq  \frac{2}{\beta_2}g\left(\frac{\langle \Delta E_1 \rangle}{\sqrt{\text{var}(\Delta E_1)}} \right)+
\left<\Delta E_1 \right>
\label{eq:Wbound-simple}
\end{align}

\begin{figure}%
    \centering
    \begin{subfigure}{}
        \includegraphics[width=7.5cm]{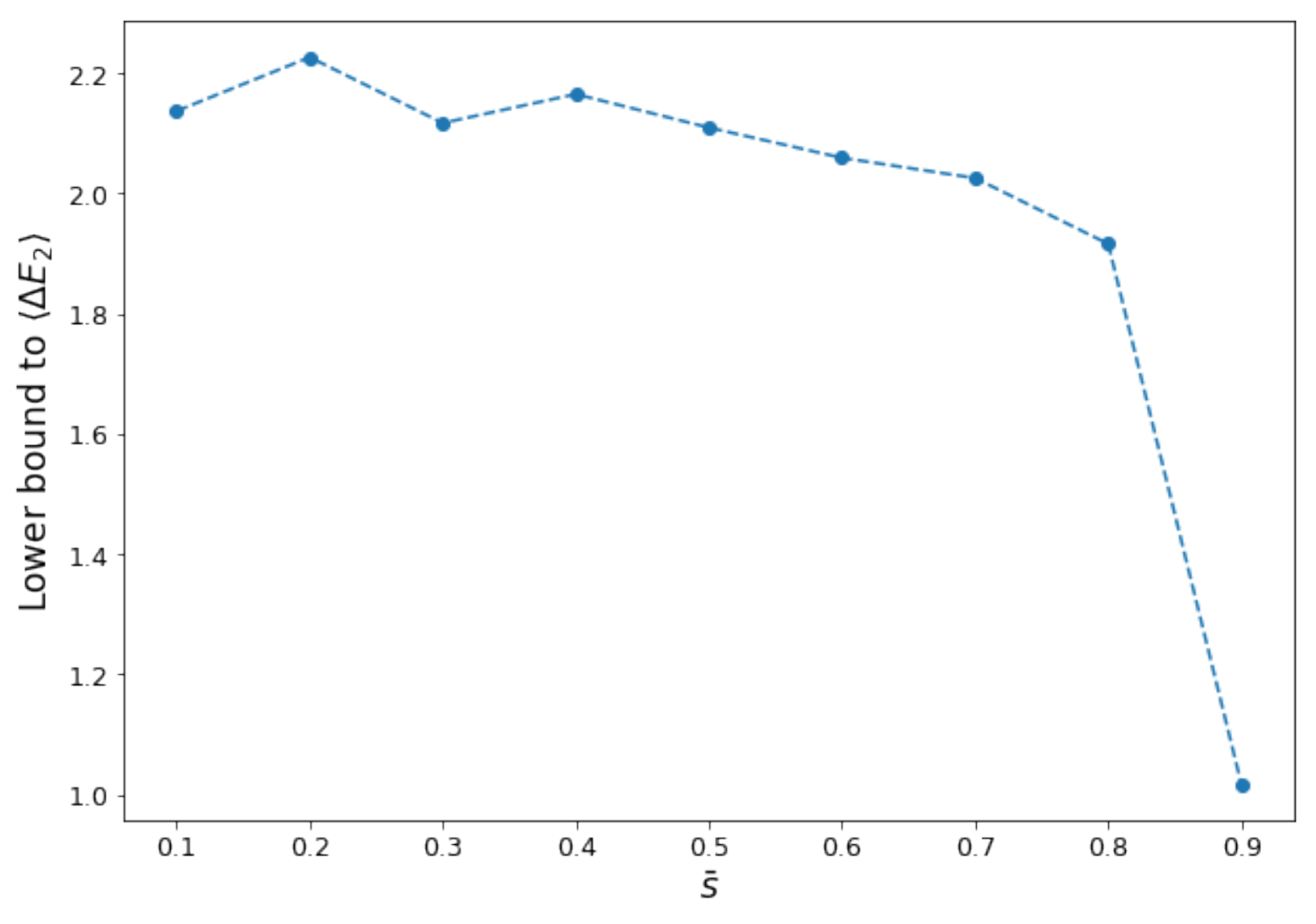}
    \end{subfigure}
    \begin{subfigure}{}
        \includegraphics[width=7.5cm]{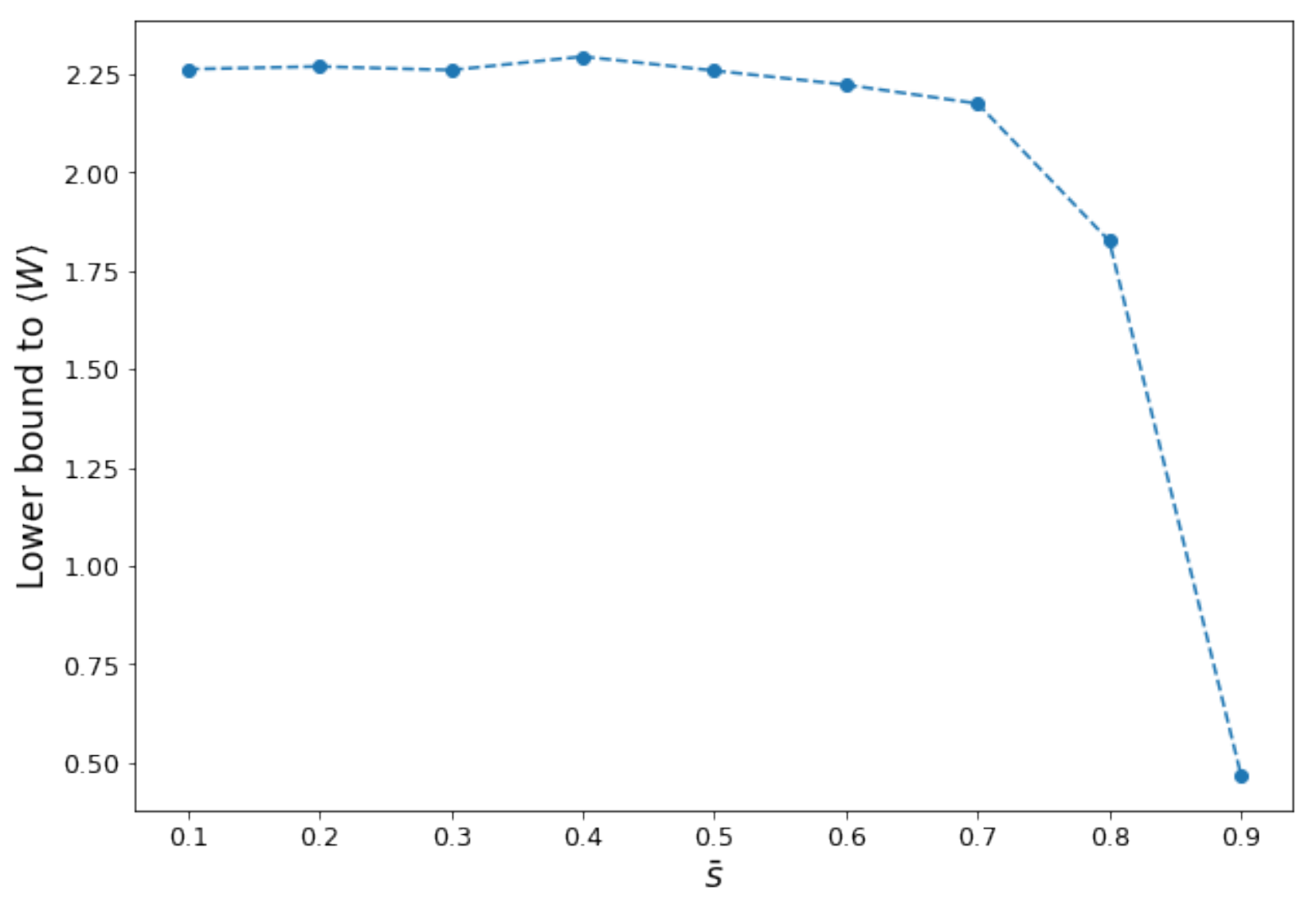}
    \end{subfigure}
    \caption{Panel a) Experimentally determined lower bound on the average energy absorbed by the cold environment $\left<\Delta E_2\right>=-Q$ as a function of $\bar{s}$, Eq. (\ref{eq:Qbound}). Panel b) Experimentally determined lower bound on the average work provided by the external driving $\left< W\right>$ as a function of $\bar{s}$, Eq. (\ref{eq:Wbound}).}
    \label{fig:work}
\end{figure}

Figure \ref{fig:work} shows the behaviour of the bounds on heat and work, as obtained from Eqs (\ref{eq:Qbound-simple}, \ref{eq:Wbound-simple}) as functions of $\bar{s}$. To achieve that, we 
estimated the temperature $T_2$ of the environment. 
To obtain such an estimation we employed the pseudo-likelikood method described in \cite{aurell2012inverse,benedetti2016estimation}. 
Given  a  set  of samples of spin configurations $\mathcal{D} = \left\lbrace s^1,...,s^D\right\rbrace$,   where $s^d= (s^d_1,...,s^d_N), d=  1,...,D$,  generated  in  a quantum  annealer  with  control parameters $J_{ij}$ and $h_i$, the estimated temperature $\hat\beta$ of their distribution is obtained by maximization of the average pseudo-likelihood
\begin{align}
    \Lambda(\beta) &= - \frac{1}{ND}\sum_{i=1}^N\sum_{d=1}^D \nonumber \\
    &\ln\left\lbrace  1+\exp\left[-2\beta s_i^d\left( h_i+\sum_{j\in\delta_i}J_{ij}s^d_j\right)\right]\right\rbrace,
\end{align}
that is
\begin{equation}
\hat \beta=\arg\max_{\beta}\Lambda(\beta)
\end{equation}
Here the symbol $\delta_i$ stands for the set of nearest neighbours of site $i$.
Using this method, we estimated $\beta_2 $ from the final processor energy distribution at various values of $\bar{s}$ and noted that for $\bar{s} \leq 0.5$ it took approximately a constant value regardless of $\bar{s}$. We then took the according value as our estimate of the environment inverse temperature. With this method we estimated $\beta_2 = 3.25$. The estimation is dependent on the specific problem we are running on the chip, and the value found is in agreement with the temperature ranges reported in \cite{benedetti2016estimation}. Note that the Hamiltonian is expressed in adimensional units, and so is $\beta_2$.

The plots discussed above evidence that $\langle \Delta E_2 \rangle>0$, and $\langle W \rangle>0$ for all values of $\bar{s}$. By inspection of Eq. (\ref{eq:modes}) we see that the only allowed operation mode having $\langle \Delta E_2 \rangle>, \langle W \rangle>0$ is the accelerator [A]. We conclude that in our experiments the D-wave operates as a thermal accelerator. Note as well that, larger energy exchanges (either in the form heat or work) occur as $\bar s$ decreases, and accordingly more dissipation occurs as evidenced above.\\

The code used to perform all the experiments as well as the aggregated data used to generate all the plots contained in this article is available to the public \cite{gitrepo}. In total we submitted $\sim 20k$ jobs to the D-Wave 2000Q processor publicly available through Leap, using a total of $\sim 5 min$ of QPU time.

\section{Conclusions}
We have observed that, as long as no external driving is applied the system slowly thermalizes, whereas, as soon as  significant amount of work is injected, thermalization speeds up dramatically with the system dumping energy to the environment. 
Accordingly our experiments evidence that, from a thermodynamic point of view, D-Wave quantum annealer behaves as a thermal accelerator. 

We deduced that, during an annealing process, the system follows the ``ground state'' not much because it evolves in agreement with the conditions of the quantum adiabatic theorem \cite{Messiah62Book}, rather because it quickly thermalises with a cold environment. In other words a quantum computation process is better understood as a cold isothermal process than a slow adiabatic process.

Our experiments also evidence that  more dissipation and larger energy exchanges are involved with larger transverse field component. This is understood on the basis of the spectral properties of the processor, whose gaps tend to become narrower as the transverse field component increases, thus opening-up the system and allowing faster thermalisation.

\section*{Acknowledgements}
We thankfully acknowledge Radomir Stevanovic and Hossein Sadeghi for the tips and help given in troubleshooting D-Wave's ocean API.

\bibliography{references} 

\end{document}